\documentclass[journal,twoside]{IEEEtran}
\usepackage{graphicx,epsfig,multirow,array}
\usepackage{bm}
\usepackage{subfigure}
\usepackage{amsmath}
\usepackage{amssymb}
\usepackage{cite}
\usepackage{stfloats}
\usepackage{multicol}
\usepackage{algorithm}
\usepackage{algorithmicx}
\usepackage{algpseudocode}
\usepackage{hhline}
\usepackage{colortbl}
\usepackage{color}

\usepackage{mathrsfs}
\usepackage{bm}
\newtheorem{thm}{Theorem}

\begin{document}

\title{Random Shifting Intelligent Reflecting Surface for OTP Encrypted Data Transmission}
\author {Zijie~Ji,
        Phee~Lep~Yeoh,~\IEEEmembership{Member,~IEEE,}
        Gaojie~Chen,~\IEEEmembership{Senior Member,~IEEE,}
        Cunhua~Pan,~\IEEEmembership{Member,~IEEE,}
        Yan~Zhang,~\IEEEmembership{Member,~IEEE,}
        Zunwen~He,~\IEEEmembership{Member,~IEEE,}
        Hao~Yin,
        and~Yonghui Li,~\IEEEmembership{Fellow,~IEEE}
\thanks{This work was supported by the National Key R\&D Program of China under Grant 2020YFB1804901, the National Natural Science Foundation of China under Grant 61871035, the China Scholarship Council scholarship, and Ericsson company. \emph{(Corresponding author: Yan Zhang.)}}
\thanks{Z.~Ji, Y.~Zhang, and Z.~He are with the School of Information and Electronics, Beijing Institute of Technology, Beijing 100081, China (e-mail: \{jizijie, zhangy, hezunwen\}@bit.edu.cn).}
\thanks{P.~L.~Yeoh and Y.~Li are with the School of Electrical and Information Engineering, University of Sydney, Sydney, NSW 2006, Australia (e-mail: \{phee.yeoh, yonghui.li\}@sydney.edu.au).}
\thanks{G.~Chen is with the Department of Engineering, University of Leicester, Leicester LE1 7RH, U.K. (e-mail: gaojie.chen@leicester.ac.uk).}
\thanks{C.~Pan is with the School of Electronic Engineering and Computer Science, Queen Mary University of London, London E1 4NS, U.K. (e-mail: c.pan@qmul.ac.uk).}
\thanks{H.~Yin is with Institute of China Electronic System Engineering Corporation, Beijing 100141, China (e-mail: yinhao@cashq.ac.cn).}
}

\markboth{IEEE WIRELESS COMMUNICATIONS LETTERS}
{Ji \MakeLowercase{\textit{et al.}}: Random Shifting Intelligent Reflecting Surface for OTP Encrypted Data Transmission}
\maketitle

\begin{abstract}
In this letter, we propose a novel encrypted data transmission scheme using an intelligent reflecting surface (IRS) to generate secret keys in wireless communication networks. We show that perfectly secure one-time pad (OTP) communications can be established by using a simple random phase shifting of the IRS elements. To maximize the secure transmission rate, we design an optimal time slot allocation algorithm for the IRS secret key generation and the encrypted data transmission phases. Moreover, a theoretical expression of the key generation rate is derived based on Poisson point process (PPP) for the practical scenario when eavesdroppers' channel state information (CSI) is unavailable. Simulation results show that employing our IRS-based scheme can significantly improve the encrypted data transmission performance for a wide-range of wireless channel gains and system parameters.
\end{abstract}

\IEEEpeerreviewmaketitle

\section{Introduction}
\IEEEPARstart{I}{ntelligent} reflecting surface (IRS) is a promising candidate technology for future mobile communication systems due to its low-cost deployment and high spectral and energy efficiency \cite{reference1,reference2}. By adaptively adjusting phase shifts of large-scale passive reflecting elements, IRS can reconfigure the electromagnetic propagation environment of wireless devices, thereby improving their communication performance \cite{reference2.1}. Recently, physical layer security in IRS assisted networks has also attracted extensive attention \cite{reference3,reference4,reference5}. However, most of these contributions focused on designing beamforming and artificial noise vectors, whilst IRS based secret key generation remains unexplored.

Physical layer secret key generation is an alternative security approach to computational complexity based encryption \cite{reference6}, where secret keys are generated by legitimate transceivers using their reciprocal time-variant channels, while preventing eavesdroppers from inferring the generated keys through its own wireless channel observations due to spatial decorrelation. Notably, the key generation rate (KGR) is limited by the movement speed of the legitimate transceivers, which makes it impractical for stationary or slow-moving wireless devices. To achieve secure one-time pad (OTP) \cite{reference7}, artificial randomness is introduced to boost the KGR. In \cite{reference8}, a method to induce randomness was proposed by selecting different local pilot constellations. In \cite{reference9}, private random precoding vectors were designed to induce randomness in multiple-input multiple-output (MIMO) systems. The authors in \cite{reference10} utilized artificial noise to scramble eavesdropping channels. We note that all the above contributions focused on increasing randomness at transceiver ends, rather than increasing the rate of channel randomness in the wireless environment.

In this letter, we consider the use of IRS to induce virtual fast fading channels, where artificial randomness is introduced in the propagation environment. Compared to the beamforming based scheme \cite{reference11} which requires high IRS hardware accuracy, we propose a simpler random phase shifting scheme for secret key generation, which only relies on the fast phase switching to support OTP encrypted data transmission. To provide detailed insights, we derive the secure transmission rate and KGR of our proposed scheme. Based on the maximum transmission rate, we develop an optimal time slot allocation algorithm for the key generation and data transmission phases. For randomly distributed eavesdroppers whose channel state information (CSI) is unknown, we further consider the use of a Poisson point process (PPP) model to derive the KGR that is only related to the distribution intensity of eavesdroppers. Finally, the advantages of our scheme and the validity of the analysis is verified through simulations. We show that the secure transmission rate is enhanced significantly by random phase shifted IRS compared with no IRS and IRS with fixed phase shifts. The impacts of important system design parameters including the length of coherence interval, the number of IRS elements, signal-to-noise ratio (SNR), the distribution intensity and distribution radius of eavesdroppers, are highlighted in the simulation results.

\section{System Model}
Our proposed system model consists of a single-antenna base station (Alice) and a single-antenna user (Bob) wanting to exchange data securely over the public wireless channels, and $K$ single-antenna eavesdroppers (Eves) attempting to capture the transmitted information. Alice and Bob exploit their reciprocal channels in time-division duplex (TDD) mode to generate keys, then encrypt and decrypt the transmitted data using the OTP approach. We assume that Alice is fixed and Bob is a slow-moving user which incurs a relative long coherence interval and thus a limited KGR. To significantly increase the KGR, we propose to use an $N$-element IRS (Rose) that can artificially adjust phase shifts and induce randomness in the wireless propagation environment. Considering Bob is a mobile user, we assume all Eves are passive and located close to Alice so that their observed channels can be correlated with the legitimate ones.

\section{Random Shifting IRS for OTP Transmission}
Fig. 1 shows our scheme which is composed of two phases: secret key generation and encrypted data transmission. Firstly, Alice and Bob exchange pilots during the first $2Q$ time slots, assisted by Rose which simply reflects the transmissions with a random set of phase shifts in each uplink and downlink round. During the odd time slots, Alice sends a pilot sequence distributed on different sub-carriers ${\bf{s}}_1 \sim \mathcal{CN}(0,{\bf{I}})$, and the received signal at Bob or the $k$th Eve can be expressed as

\begin{small}
\begin{equation}\label{equation1}
\setlength{\abovedisplayskip}{0pt}
{\bf{y}}_{i,1}^{(q)} = \sqrt P \left( {{h_{{\rm{a}}i}} + {\bf{h}}_{{\rm{r}}i}^H{{\bf{\Theta }}^{(q)}}{{\bf{h}}_{{\rm{ar}}}}} \right){{\bf{s}}_1} + {{\bf{z}}_i},\;i \in \{ {\rm{b}},{{\rm{e}}_k}\},
\end{equation}
\end{small}where ${h_{{\rm{a}}i}}\in {\mathbb{C}^{1 \times 1}}$, ${{\bf{h}}_{\rm{ar}}}\in {\mathbb{C}^{N \times 1}}$, and ${\bf{h}}_{{\rm{r}}i}\in {\mathbb{C}^{N \times 1}}$ denote the direct channel from Alice to node $i$, the incident channel from Alice to Rose, and the reflected channel from Rose to Bob or Eve, respectively. Let ${{\bf{\Theta }}^{(q)}} \!=\! {\rm{diag}}\left({e^{j{\theta _{1,q}}}},{e^{j{\theta _{2,q}}}}, \ldots ,{e^{j{\theta _{N,q}}}}\right)$ denote the diagonal IRS phase shifting matrix in the $q$th round of channel training, where $q = 1,2, \ldots ,Q$. ${{\theta _{n,q}}}$ denotes the discrete phase shift introduced by the $n$th reflecting element, which is generated based on an independent random variable (RV) uniformly distributed in a set of discrete values $\left\{ 0,\frac{{2\pi }}{{{2^B}}}{\rm{,}} \ldots {\rm{,}}\frac{{({2^B} - 1)2\pi }}{{{2^B}}} \right\}$, where $B$ is the number of quantization bits for phase shifts. In (\ref{equation1}), $P$ is the transmit power, and ${{\bf{z}}_i} \sim \mathcal{CN}(0,\sigma _i^2{\bf{I}})$ denotes the independent and identically distributed (i.i.d.) complex additive white Gaussian noise vector.

\begin{figure}[t]
  \centering
  \includegraphics[width=0.4\textwidth]{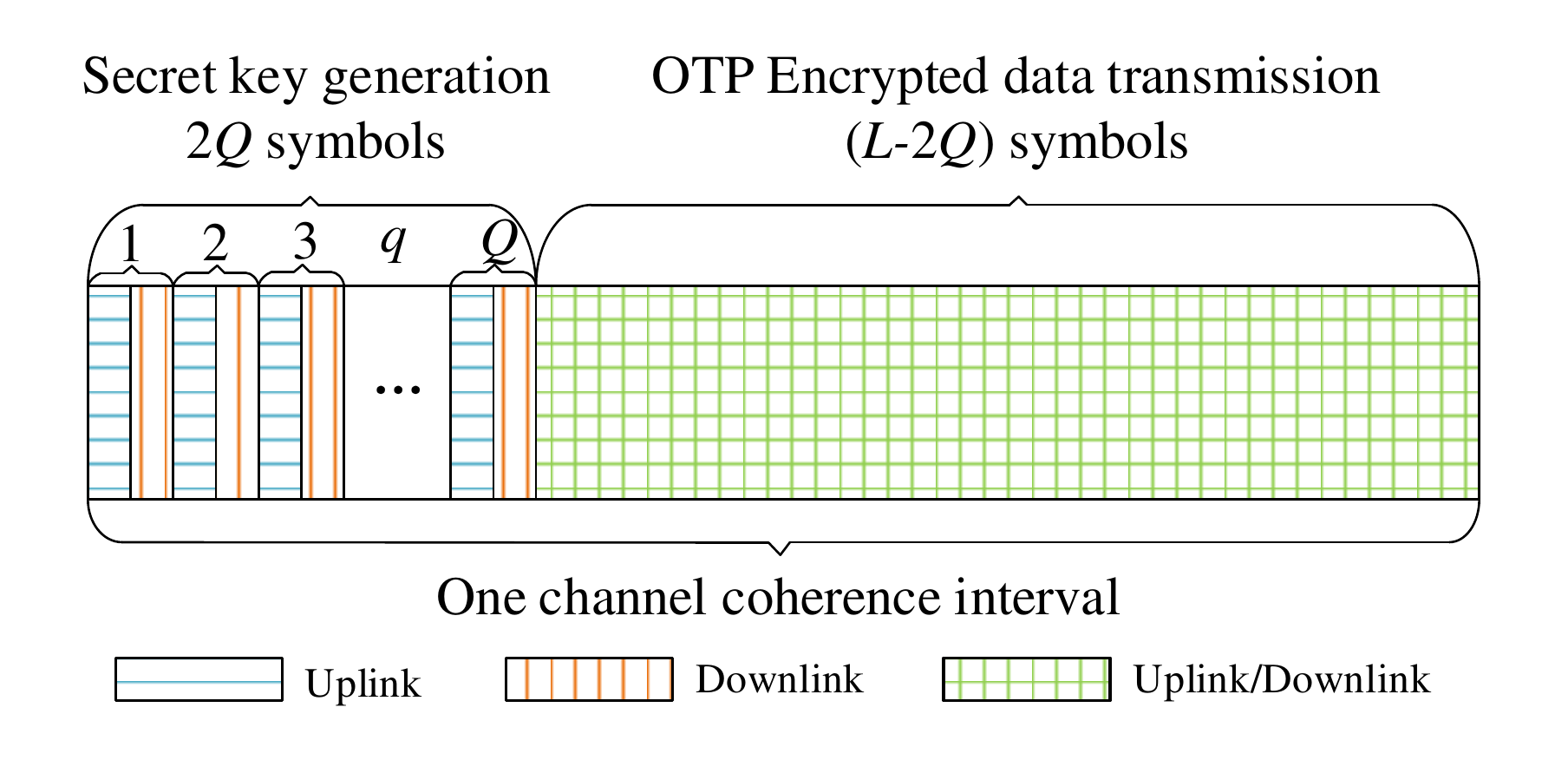}\\
  \caption{Time slot allocation for the IRS-assisted secret key generation phase and the OTP encrypted data transmission phase.}
  \label{figure1}
\end{figure}

Similarly, during the even time slots, Bob transmits ${\bf{s}}_2\sim \mathcal{CN}(0,{\bf{I}})$ with the same power, and Alice or Eve receives

\begin{small}
\begin{equation}\label{equation2}
\setlength{\abovedisplayskip}{0pt}
{\bf{y}}_{i,2}^{(q)} \!=\! \sqrt P \left({h_{{\rm{b}}i}} \!+\! {\bf{h}}_{{\rm{r}}i}^H{{\bf{\Theta }}^{(q)}}{{\bf{h}}_{{\rm{br}}}}\right){{\bf{s}}_2} \!+\! {{\bf{z}}_i},\ i \in \{ {\rm{a}},{\rm{e}}_k\},
\end{equation}
\end{small}where ${h_{{\rm{b}}i}}\in {\mathbb{C}^{1 \times 1}}$ and ${{\bf{h}}_{{\rm{br}}}}\in {\mathbb{C}^{N \times 1}}$ are the channel from Bob to node $i$ and Rose, respectively. We assume that all involved channels are quasi-static. They remain constant in each coherence interval and change independently between different coherence intervals. The IRS ensures that the equivalent channels change randomly in each round, which helps to significantly increase the KGR. The equivalent channels estimated at each node can be calculated from their observations as

\begin{small}
\begin{equation}\label{equation3}
\setlength{\abovedisplayskip}{0pt}
\begin{array}{*{20}{l}}
{h_{\rm{B}}^{(q)} \!=\! \frac{{{\bf{y}}_{{\rm{b}},1}^{(q)H}{\bf{s}}_1}}{{\sqrt P {\rm{||}}{{\bf{s}}_1}{\rm{||}}_2^2}} \!=\! g_{{\rm{ab}}}^{(q)} \!+\! {{\hat z}_{\rm{b}}} \!=\! \left({h_{{\rm{ab}}}} \!+\! {\bf{h}}_{{\rm{rb}}}^H{{\bf{\Theta }}^{(q)}}{{\bf{h}}_{{\rm{ar}}}}\right) \!+\! {{\hat z}_{\rm{b}}},}\\
{h_{\rm{A}}^{(q)} \!=\! \frac{{{\bf{y}}_{{\rm{a}},2}^{(q)H}{\bf{s}}_2}}{{\sqrt P {\rm{||}}{{\bf{s}}_2}{\rm{||}}_2^2}} \!=\! g_{{\rm{ba}}}^{(q)} \!+\! {{\hat z}_{\rm{a}}} \!=\! \left({h_{{\rm{ba}}}} \!+\! {\bf{h}}_{{\rm{ra}}}^H{{\bf{\Theta }}^{(q)}}{{\bf{h}}_{{\rm{br}}}}\right) \!+\! {{\hat z}_{\rm{a}}},}\\
{h_{{\rm{AE}}_k}^{(q)} \!=\! \frac{{{\bf{y}}_{{{\rm{e}}_k},1}^{(q)H}{\bf{s}}_1}}{{\sqrt P {\rm{||}}{{\bf{s}}_1}{\rm{||}}_2^2}} \!=\! g_{{\rm{a}}{{\rm{e}}_k}}^{(q)} \!+\! {{\hat z}_{{{\rm{e}}_k}}} \!=\! \left({h_{{\rm{a}}{{\rm{e}}_k}}} \!+\! {\bf{h}}_{{\rm{r}}{{\rm{e}}_k}}^H{{\bf{\Theta }}^{(q)}}{{\bf{h}}_{{\rm{ar}}}}\right) \!+\! {{\hat z}_{{{\rm{e}}_k}}},}\\
{h_{{\rm{BE}}_k}^{(q)} \!=\! \frac{{{\bf{y}}_{{{\rm{e}}_k},2}^{(q)H}{\bf{s}}_2}}{{\sqrt P {\rm{||}}{{\bf{s}}_2}{\rm{||}}_2^2}} \!=\! g_{{\rm{b}}{{\rm{e}}_k}}^{(q)} \!+\! {{\hat z}_{{{\rm{e}}_k}}} \!=\! \left({h_{{\rm{b}}{{\rm{e}}_k}}} \!+\! {\bf{h}}_{{\rm{r}}{{\rm{e}}_k}}^H{{\bf{\Theta }}^{(q)}}{{\bf{h}}_{{\rm{br}}}}\right) \!+\! {{\hat z}_{{{\rm{e}}_k}}},}
\end{array}
\end{equation}
\end{small}where ${\hat z}_i \sim \mathcal{CN}(0,\hat \sigma _i^2)$ denotes the estimation error at node $i$, $i \in \{ {\rm{a,b}},{{\rm{e}}_k}\}$, $\lVert\cdot\rVert_2$ is the Euclidean norm, and $\hat \sigma _i^2 = \sigma _i^2/P$.  Assume that $\sigma _{\rm{a}}^2 \!=\! \sigma _{\rm{b}}^2 \!=\! \sigma _{{{\rm{e}}_k}}^2 \!=\! \sigma^2$, and $g_{{\rm{ab}}}^{(q)}=g_{{\rm{ba}}}^{(q)}$ are the reciprocal combined channels for Alice and Bob to generate shared keys. To eliminate the impact of amplitude difference caused by hardware, the estimations are linearly normalized before quantization as $\tilde h_I^{(q)} \!=\! h_I^{(q)}/\lVert{{\bf{h}}_I}\rVert_2,\ I \!\in\! \{ {\rm{A}},{\rm{B}},{\rm{AE}}_k,{\rm{BE}}_k\}$,
where ${{{\bf{h}}}_I} \!=\! [h_I^{(1)}, h_I^{(2)}, \ldots, h_I^{(Q)}]$. Given these channels, the maximum achievable KGR can be expressed as \cite{reference6,reference9,reference12}

\begin{small}
\begin{equation}\label{equation4}
\setlength{\abovedisplayskip}{0pt}
{R_{{\rm{SKG}}}} = \mathop {\min }\limits_{k \in \mathcal K} \left( {\frac{1}{{2Q\Delta T}}\sum\limits_{q = 1}^Q {{\rm I}\left(\tilde h_{\rm{A}}^{(q)};\tilde h_{\rm{B}}^{(q)}|\tilde h_{{\rm{AE}}_k}^{(q)},\tilde h_{{\rm{BE}}_k}^{(q)}\right)} } \right),
\end{equation}
\end{small}where ${\mathcal K} \!=\! \{1,2, \ldots ,K\}$, and $\Delta T$ is the duration of one time slot. Here, we consider the average KGR per unit time.

\begin{thm}
When the number of IRS elements $N$ is sufficiently large\footnote{IRS is commonly equipped with tens or hundreds of elements to achieve high performance \cite{reference2}, so this assumption is reasonable.}, the maximum achievable KGR in our proposed scheme can be given by

\begin{small}
\begin{equation}\label{equation5}
\setlength{\abovedisplayskip}{0pt}
\begin{aligned}
{R_{{\rm{SKG}}}}  &\!=\! \mathop {\min }\limits_{k \in {\mathcal K}} (R_{{\rm{SKG}}}^{(k)})\\
&\!=\! \mathop {\min }\limits_{k \in {\mathcal K}} \left[ {\frac{1}{{2\Delta T}}{{\log }_2}\left( {\frac{{{{\left(1 \!-\! \rho _{{{\rm{E}}_k}}^2\right)}^2}}}{{1 \!+\! 2{\rho _{\rm{L}}}\rho _{{{\rm{E}}_k}}^2 \!-\! 2\rho _{{{\rm{E}}_k}}^2 \!-\! \rho _{\rm{L}}^2}}} \right)} \right],
\end{aligned}
\end{equation}
\end{small}where ${\rho _{{\rm{E}}_k}} \!=\! \mathbb{E}\{ \tilde h_{\rm{A}}^{(q)}\tilde h_{{\rm{BE}}_k}^{(q)*}\} \!=\! \mathbb{E}\{ \tilde h_{\rm{B}}^{(q)}\tilde h_{{\rm{BE}}_k}^{(q)*}\}$, ${\rho _{\rm{L}}} \!=\! \mathbb{E}\{ \tilde h_{\rm{A}}^{(q)}\tilde h_{{\rm{B}}}^{(q)*}\}$ denote the cross-correlations between $\tilde h_{\rm{A}}^{(q)}$ (or $\tilde h_{\rm{B}}^{(q)}$) and $\tilde h_{{\rm{BE}}_k}^{(q)}$, and that between $\tilde h_{\rm{A}}^{(q)}$ and $\tilde h_{\rm{B}}^{(q)}$, respectively. $\mathbb{E}\{\cdot\}$ represents the expectation with respect to (w.r.t.) the $Q$ samples.
\end{thm}

\begin{IEEEproof}
By invoking the central limit theorem, the composite channels can be approximated by the Gaussian distribution for a sufficiently large number of $N$ \cite{reference1}.
Thus, the conditional mutual information can be calculated as \cite{reference10}

\begin{small}
\begin{equation}\label{equation6}
\setlength{\abovedisplayskip}{0pt}
\begin{aligned}
&{{\rm I}\left(\tilde h_{\rm{A}}^{(q)};\tilde h_{\rm{B}}^{(q)}|\tilde h_{{\rm{AE}}_k}^{(q)},\tilde h_{{\rm{BE}}_k}^{(q)}\right)}\\
=&\: {\rm H}\left(\tilde h_{\rm{A}}^{(q)}|\tilde h_{{\rm{AE}}_k}^{(q)},\tilde h_{{\rm{BE}}_k}^{(q)}\right) - {\rm H}\left(\tilde h_{\rm{A}}^{(q)}|\tilde h_{\rm{B}}^{(q)},\tilde h_{{\rm{AE}}_k}^{(q)},\tilde h_{{\rm{BE}}_k}^{(q)}\right)\\
=&\: {\log _2}\frac{{\det \left({{\bf{W}}_{{\rm{AAE}}_k{\rm{BE}}_k}}\right)\det \left({{\bf{W}}_{{\rm{BAE}}_k{\rm{BE}}_k}}\right)}}{{\det \left({{\bf{W}}_{{\rm{AE}}_k{\rm{BE}}_k}}\right)\det \left({{\bf{W}}_{{\rm{ABAE}}_k{\rm{BE}}_k}}\right)}},
\end{aligned}
\end{equation}
\end{small}where $\det\left(\cdot\right)$ is the matrix determinant, and

\begin{small}
\begin{equation}\label{equation7}
\setlength{\abovedisplayskip}{0pt}
\begin{aligned}
{{\bf{W}}_{{\rm{AAE}}_k{\rm{BE}}_k}} &= \mathbb{E}\left\{ {\left( {\begin{array}{*{20}{c}}
{\tilde h_{\rm{A}}^{(q)}}\\
{\tilde h_{{\rm{AE}}_k}^{(q)}}\\
{\tilde h_{{\rm{BE}}_k}^{(q)}}
\end{array}} \right)\left( {\tilde h_{\rm{A}}^{(q)*}\;\tilde h_{{\rm{AE}}_k}^{(q)*}\;\tilde h_{{\rm{BE}}_k}^{(q)*}} \right)} \right\}\\
&= \left[ {\begin{array}{*{20}{l}}
{{K_{{\rm{AA}}}}}&{{K_{{\rm{AAE}}_k}}}&{{K_{{\rm{ABE}}_k}}}\\
{{K_{{\rm{AE}}_k{\rm{A}}}}}&{{K_{{\rm{AE}}_k{\rm{AE}}_k}}}&{{K_{{\rm{AE}}_k{\rm{BE}}_k}}}\\
{{K_{{\rm{BE}}_k{\rm{A}}}}}&{{K_{{\rm{BE}}_k{\rm{AE}}_k}}}&{{K_{{\rm{BE}}_k{\rm{BE}}_k}}}
\end{array}} \right].
\end{aligned}
\end{equation}
\end{small}${K_{IJ}} \!=\! \mathbb{E}\{\tilde h_I^{(q)}\tilde h_J^{(q)*}\}, I,J \!\in\! \{ {\rm{A}},{\rm{B}},{\rm{AE}}_k,{\rm{BE}}_k\}$ is the correlation function. When two nodes are more than half-wavelength apart, their observed channels can be deemed uncorrelated \cite{reference8}. Hence, here we only consider ${\rho _{{\rm{E}}_k}}$ and ${\rho _{\rm{L}}}$ for simplicity. Since the channel observations of two nodes separated by more than half of the wavelength can be considered uncorrelated \cite{reference8}, \cite{reference12} and all channels have been normalized, we have

\begin{small}
\begin{equation}\label{equation8}
\setlength{\abovedisplayskip}{0pt}
\det \left({{\bf{W}}_{{\rm{AAE}}_k{\rm{BE}}_k}}\right) = \det \left( {\begin{array}{*{20}{c}}
1&0&{\rho _{{\rm{E}}_k}}\\
0&1&0\\
{\rho _{{\rm{E}}_k}}&0&1
\end{array}} \right) = 1 - \rho _{{\rm{E}}_k}^2.
\end{equation}
\end{small}Similarly, the determinants of other matrices in (\ref{equation6}) can be given as $\det \left({{\bf{W}}_{{\rm{BAE}}_k{\rm{BE}}_k}}\right) \!=\! 1 \!-\! \rho _{{{\rm{E}}_k}}^2$, $\det \left({{\bf{W}}_{{\rm{AE}}_k{\rm{BE}}_k}}\right) \!=\! 1$, and $\det \left({{\bf{W}}_{{\rm{ABAE}}_k{\rm{BE}}_k}}\right) \!=\! 1 + 2{\rho _{\rm{L}}}\rho _{{{\rm{E}}_k}}^2 - 2\rho _{{{\rm{E}}_k}}^2 - \rho _{\rm{L}}^2$. In addition, since the phase shift coefficients are randomly selected, the autocorrelation between two channels sampled in different rounds denoted as $\tilde h_{I}^{({q_1})}$ and $\tilde h_{I}^{({q_2})}$ is expressed as

\begin{small}
\begin{equation}\label{equation9}
\setlength{\abovedisplayskip}{0pt}
\begin{aligned}
\rho _I^{({q_1},{q_2})}
&\!=\! \frac{{\sigma _{{h_{ij}}}^2 \!+\! \sigma _{{h_{{\rm{r}}j}}}^2\sigma _{{h_{i{\rm{r}}}}}^2\sum\nolimits_{n = 1}^N \mathbb{E}{\{ {e^{j{\theta _{n,{q_1}}}}}\} \mathbb{E}\{ {e^{ - j{\theta _{n,{q_2}}}}}\} } }}{{\sigma _{{h_{ij}}}^2 \!+\! N\sigma _{{h_{{\rm{r}}j}}}^2\sigma _{{h_{i{\rm{r}}}}}^2}},
\end{aligned}
\end{equation}
\end{small}where ${\sigma _{h_{ij}}^2}$, ${\sigma _{h_{{\rm{r}}j}}^2}$, and ${\sigma _{{h_{i{\rm{r}}}}}^2}$ are the average power of ${h_{ij}}$, ${h_{{{\rm{r}}j,}n}}$, and ${h_{{i{\rm{r}},}n}}$ (the $n$th element of ${{\bf{h}}_{{\rm{r}}j}}$ and ${{\bf{h}}_{i{\rm{r}}}}$), $i,j \in \{ {\rm{a,b,}}{{\rm{e}}_k}\}$, respectively. Since ${\theta _{n,q}} \sim {\mathcal U}(0,2\pi )$, $\mathbb{E}\{ {e^{j{\theta _{n,q}}}}\}=0$. If $N \gg \sigma _{{h_{ij}}}^2/(\sigma _{{h_{{\rm{r}}j}}}^2\sigma _{{h_{i{\rm{r}}}}}^2)$, it shows that $\rho _{I}^{({q_1},{q_2})} \to 0$. This condition always holds under our assumption on $N$, especially when the IRS reflected channels are strong. Thus, the channel samples in different rounds can be deemed as independent RVs, and the maximum achievable KGR can be computed by (\ref{equation5}) based on the weak law of large numbers.
\end{IEEEproof}

Based on (\ref{equation5}), Alice and Bob can determine relevant parameters and generate keys. We note that standard key generation processing such as equally likely quantization, reconciliation and privacy amplification should be considered as in \cite{reference6}.

After key generation, the data to be transmitted is encrypted by XORing with the generated keys using the OTP approach, and then we consider that the ciphertext is transmitted at the maximum rate for the IRS-assisted system using maximum rate transmission (MRT) as considered in \cite{reference1}. We note that since the keys have been shared between Alice and Bob, this encrypted data transmission stage can be either uplink or downlink. Taking the downlink for an example, the transmission rate is given by

\begin{small}
\begin{equation}\label{equation10}
\setlength{\abovedisplayskip}{0pt}
{R_{{\rm{MRT}}}} = \frac{1}{{\Delta T}}{\log _2}\left( {1 + \gamma_{\rm b} |{{h}_{{\rm{ab}}}} + {{{\bf{h}}}_{{\rm{rb}}}^H}{{\bf{\Theta }}^ \star }{{{\bf{h}}}_{{\rm{ar}}}}{|^2}} \right),
\end{equation}
\end{small}where $\gamma_{\rm b}  = {P/{\sigma _{\rm b}^2}}$ is the reference SNR and ${\bf{\Theta }}^\star$ is the optimal IRS phase shift matrix defined in [1, (19)]. Let $L$ denote the total number of symbols in each coherence interval, thus the remaining $\left(L-2Q\right)$ time slots is used for encrypted data transmission. We observe that there exists a tradeoff between the two phases: if $Q$ is too small, the generated keys will be insufficient to encrypt all the data to be transmitted; whereas if $Q$ is too large, the remaining time slots for data transmission will be insufficient and thus some generated keys will be wasted. Therefore, the secure transmission throughput in our proposed scheme depends on the minimum of the generated key bits and the transmitted data bits, where the secure transmission rate is derived as

\begin{small}
\begin{equation}\label{equation11}
\setlength{\abovedisplayskip}{0pt}
{R_{{\rm{EDT}}}} =
  \begin{cases}
    {R_{{\rm{SKG}}}}Q/L, &\text{if $\alpha  \le \left(L-2Q\right)/Q$},\\
	{R_{{\rm{MRT}}}}\left(L-2Q\right)/L, &\text{otherwise},
  \end{cases}
\end{equation}
\end{small}where $\alpha  = {R_{{\rm{SKG}}}}/{R_{{\rm{MRT}}}}$ is the rate ratio. To achieve the highest secure transmission rate, $Q$ should be optimized so that the length of generated keys and the length of transmitted data are closest to each other, i.e., ${Q^ \star} = \mathop {\arg \min }\limits_{Q \in {\mathcal Q}} |Q{R_{{\rm{SKG}}}} - \left(L-2Q\right){R_{{\rm{MRT}}}}|$, where ${\mathcal Q} \!=\! \{Q_{\rm th}, \ldots ,\left\lceil L/2 \right\rceil\!-\!1\}$ is the set of $Q$, and $Q_{\rm th}$ is the threshold to ensure a sufficient number of samples\footnote{Although infinite samples are required to achieve ${R_{{\rm{SKG}}}}$ theoretically \cite{reference13}, in practice we only need to ensure the number of samples is sufficiently large to obtain reliable statistics according to the weak law of large numbers.}, and thereby the capacity of the proposed scheme can be expressed as ${C_{{\rm{EDT}}}} = {R_{{\rm{EDT}}}}({Q^\star})$. In practice, because $Q$ is an integer, given the expression of ${R_{{\rm{SKG}}}}$ and ${R_{{\rm{MRT}}}}$, we can find $Q^\star$ by using the optimization approach in Algorithm~\ref{algorithm1}. All channel knowledge related to calculation is acquired and processed at the IRS end as described in Step 1 of Algorithm~\ref{algorithm1}, while Alice and Bob only need to generate keys based on their channel observations and the value of $Q^\star$ sent from the IRS.

\begin{algorithm}[tb]
\caption{Proposed Optimal Time Slot Allocation}
\label{algorithm1}
\begin{algorithmic}[1]
\State Let Rose be in the receiving mode. Alice and Bob alternatively send pilot signals so that Rose can estimate ${{\bf{h}}}_{{\rm{ar}}}$ and ${{{\bf{h}}}_{{\rm{rb}}}^H}$. Alice estimates $h_{\rm {ab}}$ and sends its phase to Rose, then ${\bf{\Theta }}^\star$ and $R_{\rm {MRT}}$ can be calculated based on (\ref{equation10}).
\State Initialize $Q^{(t)}\!=\!Q_{\rm th}$.
\Repeat \;{(Observation accumulation for ${R_{{\rm{SKG}}}^{(t)}}$)}
\State For given $Q^{(t)}$, let Rose be in the reflecting mode with randomly selected ${{\bf{\Theta }}^{(q)}}$. Alice and Bob collect channel observations $h_{\rm{A}}^{(q)}, h_{\rm{B}}^{(q)},$ and $h_{{\rm{BE}}_k}^{(q)}$ and normalize them. The corresponding ${R_{{\rm{SKG}}}^{(t)}}$ can be computed according to (\ref{equation5}).
\State Initialize $Q_{\rm min}= Q^{(t)}$ and $Q_{\rm max}= \left\lceil L/2 \right\rceil\!-\!1$.
\Repeat \;{(Bisection search for $Q^ {(t)\star}$)}
\State Set $Q\!=\!\left\lfloor {(Q_{\rm min}\!+\!Q_{\rm max})/2} \right\rfloor$, and calculate $\Delta R = Q{R_{{\rm{SKG}}}^{(t)}} - (L - 2Q){R_{{\rm{MRT}}}}$. If $\Delta R \le 0$, update $Q_{\rm min}=Q$; otherwise update $Q_{\rm max}=Q$.
\Until $(Q_{\rm max}-Q_{\rm min}) \le 1$. Record this $Q$ as $Q^ {(t)\star}$.
\State Update $Q^{(t)}=Q^{(t)}+1$.
\Until $Q^{(t)} \!\ge\! Q^ {(t-1)\star}$. Finally, we obtain $Q^ \star\!=\!Q^ {(t-1)\star}$.
\end{algorithmic}
\end{algorithm}

\section{Performance Analysis for Randomly Distributed Eavesdroppers}
We find that $R_{\rm{SKG}}$ in (\ref{equation5}) depends on the minimum $R_{{\rm{SKG}}}^{(k)}$, which is computed based on knowledge of $\tilde h_{\rm{A}}^{(q)}$, $\tilde h_{{\rm{B}}}^{(q)}$, and $\tilde h_{{\rm{BE}}_k}^{(q)}$, where $\tilde h_{{\rm{BE}}_k}^{(q)}$ is typically assumed to be known at Alice and Bob when the untrusted nodes are other users in the same network. However, we note that the CSI of passive and malicious Eves may be challenging to estimate and may not be perfectly known at Alice and Bob. To this end, in this section we focus on the performance analysis of our proposed scheme when there exists randomly distributed Eves.

Since the transmitted data has been XORed by OTP keys, Eves will not impact on the encrypted transmission stage. Only key generation can be compromised in the presence of Eves.

From (\ref{equation5}), we can derive that

\begin{small}
\begin{equation}\label{equation12}
\setlength{\abovedisplayskip}{0pt}
\begin{aligned}
\frac{{\partial R_{{\rm{SKG}}}^{(k)}}}{{\partial \rho _{{{\rm{E}}_k}}^2}} &\!=\! \frac{{\ln 2\left(\rho _{{{\rm{E}}_k}}^2 - {\rho _{\rm{L}}}\right)\left(1 - \rho _{{{\rm{E}}_k}}^2\right)}}{{\Delta T \left(1 + {\rho _{\rm{L}}} - 2\rho _{{{\rm{E}}_k}}^2\right)}} < 0,\\
\frac{{\partial R_{{\rm{SKG}}}^{(k)}}}{{\partial {\rho _{\rm{L}}}}} &\!=\! \frac{{\ln 2\left({\rho _{\rm{L}}} - \rho _{{{\rm{E}}_k}}^2\right)}}{{\Delta T\left(1 - {\rho _{\rm{L}}}\right)\left(1 + {\rho _{\rm{L}}} - 2\rho _{{{\rm{E}}_k}}^2\right)}} > 0,
\end{aligned}
\end{equation}
\end{small}when ${\rho _{\rm{L}}} > \rho _{{{\rm{E}}_k}}^2$ holds. This condition is always satisfied because noise is the only factor that results in the decrease of ${\rho _{\rm{L}}} \!=\! \mathbb{E}\{ |g_{{\rm{ab}}}^{(q)}{|^2}\} /\left(\mathbb{E}\{ |g_{{\rm{ab}}}^{(q)}{|^2}\}  \!+\! {{\hat \sigma }^2}\right)$, which shows the high similarity between observations $\tilde h_{\rm{A}}^{(q)}$ and $\tilde h_{{\rm{B}}}^{(q)}$ of the reciprocal channel ${g_{{\rm{ab}}}^{(q)}}$, but ${\rho _{{\rm{E}}_k}}$ is weakened by the difference between the combined channels ${g_{{\rm{ab}}}^{(q)}}$ and ${g_{{\rm{b}}{{\rm{e}}_k}}^{(q)}}$ in addition to noise. Hence, $R_{{\rm{SKG}}}^{(k)}$ decreases as $\rho_{{\rm E}_k}$ increases or $\rho_{\rm L}$ decreases.

For a given SNR, $\rho_{\rm L}$ is fixed, while $\rho_{{\rm E}_k}$ is affected by the positions of Eves. Since Eves are passive, we assume that their exact locations are not known. Therefore, we consider a PPP to model the locations of randomly distributed Eves and analyze the impact on $R_{{\rm{SKG}}}$. Generally, a closer distance between Alice and Eve $k$, $d_{{\rm AE}_k}$, is associated with a stronger correlation ${\rho _{{\rm{E}}_k}}$, so we need to find the expectation of the closest distance $d_{\rm min}$, which corresponds to the largest $\rho_{{\rm E}_k}$ and $R_{{\rm{SKG}}}$. Specifically, we model the locations of Eves as a homogenous PPP in the plane denoted by ${\Phi _{\rm{E}}}$ with intensity $\lambda _{\rm{E}}$. The probability density function (PDF) of $d_{{\rm AE}_k}$ of the $k$th nearest Eve to Alice is expressed as \cite{reference14}

\begin{small}
\begin{equation}\label{equation13}
\setlength{\abovedisplayskip}{0pt}
{f_{{d_{{\rm{A}}{{\rm{E}}_k}}}}}({d_{{\rm{A}}{{\rm{E}}_k}}}) = {e^{ - {\lambda _{\rm{E}}}\pi d_{{\rm{A}}{{\rm{E}}_k}}^2}}\frac{{2\lambda _{\rm{E}}^k{\pi ^k}d_{{\rm{A}}{{\rm{E}}_k}}^{2k - 1}}}{{\Gamma (k)}},
\end{equation}
\end{small}where $\Gamma (\cdot)$ is the gamma function. As such, the expectation of $d_{\rm min}$ (i.e., $d_{{\rm{A}}{{\rm{E}}_1}}$) can be calculated as

\begin{small}
\begin{equation}\label{equation14}
\setlength{\abovedisplayskip}{0pt}
\begin{aligned}
\mathbb{E}_{{\Phi _{\rm{E}}}}({d_{{\rm{min}}}})
&\!=\! \int_0^\infty  {2{\lambda _{\rm{E}}}\pi d_{{\rm{min}}}^2}{e^{ - {\lambda _{\rm{E}}}\pi d_{{\rm{min}}}^2}}{\rm d}{d_{{\rm{min}}}}\\
&\!=\! {\left( {\int_0^{\frac{\pi }{2}} {\int_0^\infty  {r{e^{ - {\lambda _{\rm{E}}}\pi {r^2}}}{\rm d}\theta {\rm d}r} } } \right)^{\frac{1}{2}}} \!=\! \sqrt {\frac{1}{{4{\lambda _{\rm{E}}}}}},
\end{aligned}
\end{equation}
\end{small}which indicates that when $\lambda _{\rm{E}}$ is high, ${d_{{\rm{min}}}}$ is expected to be small. Furthermore, the correlation between ${g_{{\rm{ab}}}^{(q)}}$ and ${g_{{\rm{b}}{{\rm{e}}_k}}^{(q)}}$ can be characterized by \cite{reference8}

\begin{small}
\begin{equation}\label{equation15}
\setlength{\abovedisplayskip}{0pt}
\tilde \rho ({d_{{\rm{A}}{{\rm{E}}_k}}}) = {[{J_0} (2\pi {d_{{\rm{A}}{{\rm{E}}_k}}}/\lambda )]^2},
\end{equation}
\end{small}where $J_0 (\cdot)$ is the Bessel function of the first kind, and $\lambda$ is the wavelength. Thus, the KGR in (\ref{equation5}) can be simplified as

\begin{small}
\begin{equation}\label{equation16}
\setlength{\abovedisplayskip}{0pt}
{R_{{\rm{SKG}}}} = \frac{1}{{2\Delta T}}{\log _2}\left( {\frac{{{{\left(1 \!-\! \rho _{\rm E\max }^2\right)}^2}}}{{1 \!+\! 2{\rho _{\rm{L}}}\rho _{\rm E\max }^2 \!-\! 2\rho _{\rm E\max }^2 \!-\! \rho _{\rm{L}}^2}}} \right),
\end{equation}
\end{small}where the maximum correlation among all ${\rho _{{\rm{E}}_k}}$ is

\begin{small}
\begin{equation}\label{equation17}
\setlength{\abovedisplayskip}{0pt}
{\rho _{\rm E\max }} = \frac{{\mathbb{E}\{ |g_{{\rm{ab}}}^{(q)}||g_{{\rm{b}}{{\rm{e}}_1}}^{(q)}|\} \tilde \rho \left(\sqrt {1/4{\lambda _{\rm{E}}}} \right)}}{{\sqrt {\mathbb{E}\{ |g_{{\rm{ab}}}^{(q)}{|^2}\}  + {{\hat \sigma }^2}} \sqrt {\mathbb{E}\{ |g_{{\rm{b}}{{\rm{e}}_1}}^{(q)}{|^2}\}  + {{\hat \sigma }^2}} }}.
\end{equation}
\end{small}With this closed-form expression, we can investigate the influence of randomly distributed Eves on the security performance of our proposed scheme. Correspondingly, $h_{{\rm{BE}}_k}^{(q)}$ in Step 4 of Algorithm~\ref{algorithm1} is no longer required, and Alice and Bob only need to determine a suitable ${\lambda _{\rm{E}}}$ according to their estimation of the surrounding network conditions.

\section{Simulation Results and Discussion}
In this section, simulation results are presented to show the effectiveness of the proposed scheme by comparing with two benchmarks: (1) without IRS, (2) IRS with fixed optimal phases. We assume that the distance between Alice and Bob is $d_{\rm {AB}}=100$ m. The central point of the IRS is located at a vertical distance of $d_1=5$ m to the line that connects Alice and Bob, and the horizontal distance between the IRS and Bob is set as $d_2=5$ m. All Eves are randomly deployed within a circle of radius 1 m centered at Alice and their distances to Bob and Rose can be calculated correspondingly. The path loss is given by $PL \!=\! \left(PL_0 \!+\! 10\zeta{\log _{10}}(d/{d_0})\right)$ dB, where $PL_0 \!=\! 30$ dB is the path loss at reference distance $d_0 \!=\! 1$ m,  $\zeta$ is the path loss exponent, and $d$ is the distance between the transmitter and the receiver. The path loss exponents for the Alice-Rose link (Eve-Rose link), Rose-Bob link, and Alice-Bob link (Eve-Bob link) are ${\zeta _{{\rm{AR}}}} = {\zeta _{{\rm{ER}}}} = 2.2$, ${\zeta _{{\rm{RB}}}} = 2.5$, and ${\zeta _{{\rm{AB}}}} = {\zeta _{{\rm{EB}}}} = 3.5$, respectively. The small-scale fading of all involved channels follows a Rayleigh fading model, and the correlation between channel coefficients of legitimate channels and eavesdropping channels is characterized by (\ref{equation15}). Other parameters are set as: the carrier frequency $f_{\rm c} \!=\! 1$ GHz, the transmit power $P \!=\! 20$ dBm, the noise power $\sigma ^2 \!=\! -96$ dBm, $\Delta T \!=\! 1$ ms, $L \!=\! 1000$, $Q_{\rm th}$ \!=\! 100, $N \!=\! 50$, $B \!=\! 3$, and $K \!=\! 4$ if not specified otherwise.

Fig. 2 shows the advantages of the proposed scheme over other benchmarks. IRS with fixed optimal phases yields higher $R_{{\rm{SKG}}}$ and $R_{{\rm{MRT}}}$ by improving the SNR compared with the case without IRS, but only one set of keys can be generated by using the existing channels. However, in our scheme, since the random IRS shifting is employed, the combined channels leads to $Q^ \star$ times of key generation and therefore $C_{{\rm{EDT}}}$ is increased. In addition, we see that the performance is significantly improved when the transmit power $P$ is increased since the corresponding high SNR benefits to both key generation and encrypted data transmission. The larger number of elements $N$ also results in better performance, but the gain brought by the same number of IRS elements will gradually decrease as $N$ increases. It can be observed that the number of quantization bits for phase shifts $B$ basically does not affect $C_{{\rm{EDT}}}$, which indicates that our scheme can still provide good performance when equipping low-resolution hardware at the IRS.

\begin{figure}[t]
  \centering
  \includegraphics[width=0.4\textwidth]{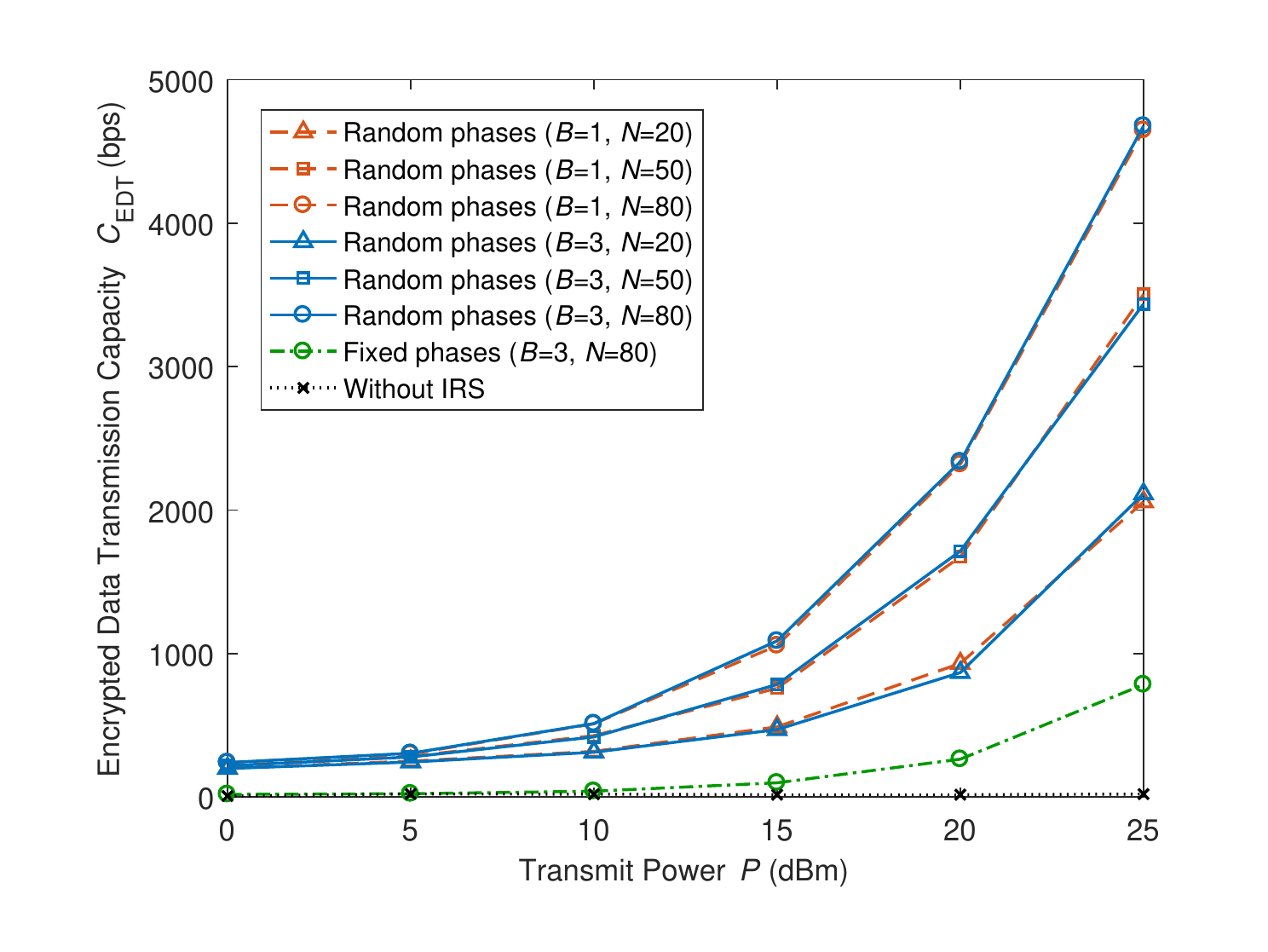}\\
  \caption{Encrypted transmission capacity $C_{\rm {EDT}}$ versus transmit power $P$ for different schemes, number of quantization bits $B$ and number of elements $N$.}
  \label{figure2}
\end{figure}

Fig. 3 depicts the impact of time slot allocation on $R_{{\rm{EDT}}}$ as discussed in Section III. At first, as more keys are generated, $R_{{\rm{EDT}}}$ increases with $Q$, then as $Q$ further increases, $R_{{\rm{EDT}}}$ decreases due to insufficient time slots assigned for encrypted data transmission. The $Q$ corresponding to each peak is $Q^ \star$, which can be found by applying Algorithm~\ref{algorithm1}, and the value of each peak pointed by the arrow is $C_{{\rm{EDT}}}$. Under the same $P$, $Q^ \star/L$ basically remains unchanged and the slight performance improvement comes from the increased accuracy of allocation proportion. This demonstrates that our scheme can be applied to various scenarios with a wide range of coherence intervals $L$. When $P$ decreases, the achievable $R_{{\rm{EDT}}}$ reduces as well, which is consistent with the analysis in Fig. 2. Meanwhile, we note that $Q^ \star$ becomes larger because the SNR impacts more on secret key generation, which compromises reciprocity between legitimate channels and thereby decreases ${\rho _{\rm{L}}}$ and $R_{{\rm{SKG}}}$.

\begin{figure}[t]
  \centering
  \includegraphics[width=0.4\textwidth]{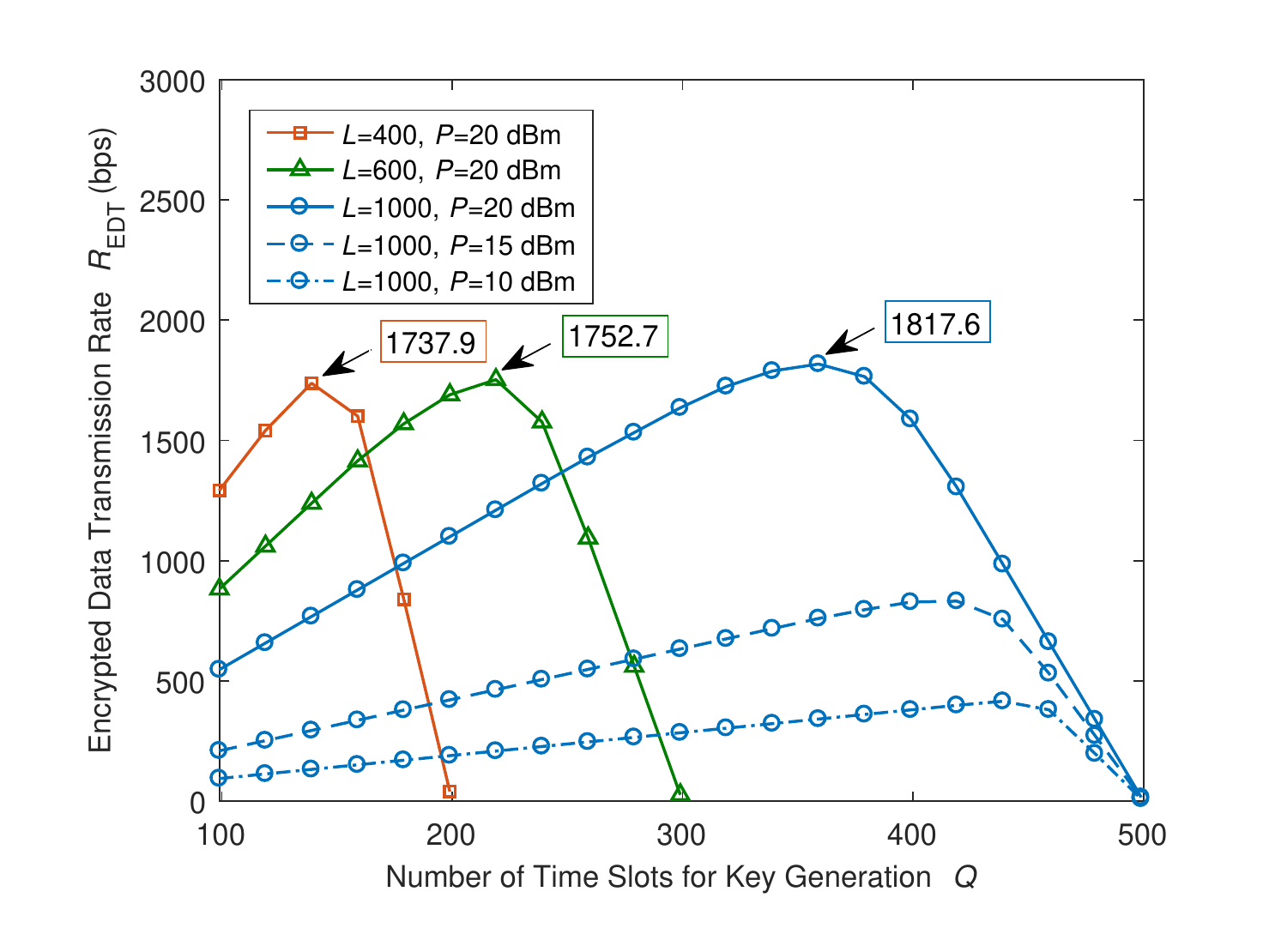}\\
  \caption{The impact of time slot allocation $Q$ on encrypted transmission rate $R_{\rm {EDT}}$ for different coherence interval $L$ and transmit power $P$.}
  \label{figure3}
\end{figure}

Fig. 4 investigates the impact of randomly distributed Eves and their distribution parameters. We can see that no matter how Eves are distributed, $R_{\rm {SKG}}$ almost increases linearly with the number of IRS elements $N$. Moreover, when the radius of ${\Phi _{\rm{E}}}$ is $R=0.1$ m, a lower $\lambda_{\rm E}$ (less Eves) also improves $R_{\rm {SKG}}$, which is consistent with our analysis in Section IV. However, this improvement becomes negligible when $R$ is increased to 1 m. This confirms the conclusion that Eves beyond several wavelengths can be ignored in secret key generation. Finally,
simulation results given by (\ref{equation5}) coincide well with theoretical results obtained from (\ref{equation16}), which shows the correctness of our analysis.

\begin{figure}[t]
  \centering
  \includegraphics[width=0.4\textwidth]{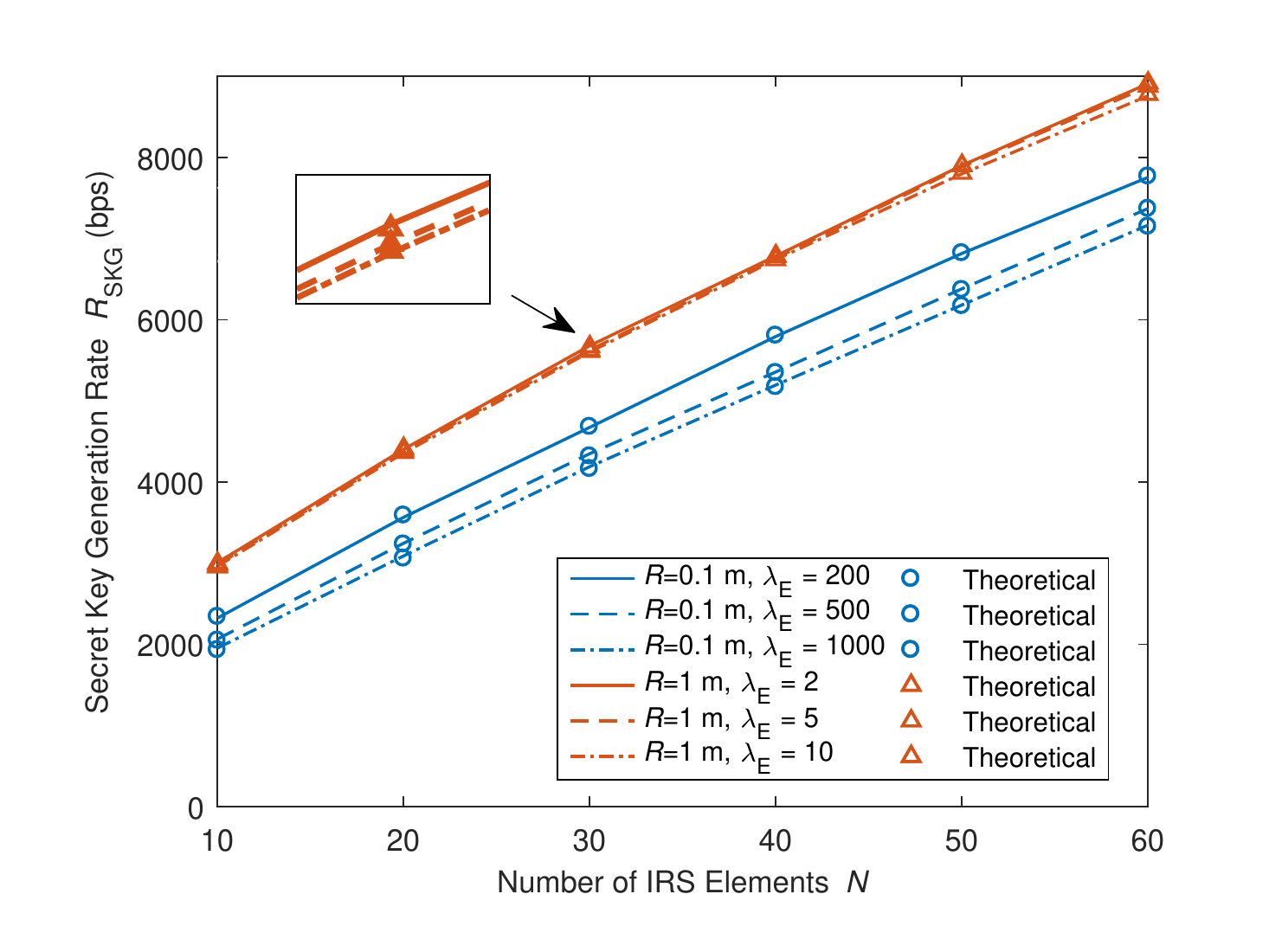}\\
  \caption{The impact of randomly distributed Eves on secret key generation rate $R_{\rm {SKG}}$ for different distribution radius $R$ and intensity $\lambda _{\rm{E}}$.}
  \label{figure4}
\end{figure}

We note that, similar to previous works such as \cite{reference1} and \cite{reference4}, the performance improvement is sensitive to the placement of the IRS. In our proposed scheme, the IRS should be deployed near the transceivers since the large-scale fading effect due to increasing distance will increase the autocorrelation in (9) thus impair the randomness of keys when the IRS is far away.

\section{Conclusions}
We investigated the confidential transmission in IRS assisted wireless networks with multiple passive eavesdroppers. A new encrypted data transmission scheme was designed, where the OTP secret keys were provided by random IRS phase shifting. The KGR was derived based on the assumption that all Eves were located near Alice, and an optimal time slot allocation algorithm was proposed to maximize the secure transmission rate. For practical implementations, we further analyzed the impact of randomly distributed Eves whose CSI is unavailable. The effectiveness of the proposed scheme and correctness of our analysis were validated by the simulation and theoretical results.

\end{document}